\documentclass[aps,prl,10pt,twocolumn,longbibliography,superscriptaddress]{revtex4-1}

\usepackage{amsmath}
\usepackage{amsthm}
\usepackage{latexsym}
\usepackage{amssymb}
\usepackage{amsfonts}
\usepackage{bm}
\usepackage{bbm}
\usepackage{appendix}
\usepackage{graphicx}
\usepackage{epstopdf}
\usepackage{mathrsfs}
\usepackage[thinc]{esdiff}
\usepackage{float}       
\usepackage{placeins}
\usepackage{color}
\usepackage[normalem]{ulem}
\usepackage{caption}
\usepackage{subcaption}
\usepackage{times}

\DeclareMathOperator{\sech}{sech}
\DeclareMathOperator{\csch}{csch}

\newcommand{\expval}[1]{\langle#1\rangle}
\newcommand{\ket}[1]{\lvert#1\rangle}
\newcommand{\bra}[1]{\langle#1\rvert}
\newcommand{\braket}[2]{\langle#1\!\mid\!#2\rangle}

\usepackage[colorlinks=true,linkcolor=blue,citecolor=green,plainpages=false,pdfpagelabels]{hyperref}
\begin{document}

\title{Relativistic Quantum-Speed Limit for Gaussian Systems and Prospective Experimental Verification}

\author{Salman Sajad Wani}
\affiliation{ Qatar Center for Quantum Computing, Hamad Bin Khalifa University, Doha, Qatar}

\author{Aatif Kaisar Khan}
\affiliation{ Department of Physics, University of Florida, Gainesville, FL 32611-8440, USA}

\author{Saif Al-Kuwari}
\affiliation{ Qatar Center for Quantum Computing, Hamad Bin Khalifa University, Doha, Qatar}

\author{Mir Faizal}
\affiliation{ Canadian Quantum Research Center, 204-3002 32 Ave Vernon, BC V1T 2L7 Canada}
\affiliation{ Irving K. Barber School of Arts and Sciences, University of British Columbia - Okanagan, Kelowna, British Columbia V1V 1V7, Canada}
\affiliation{ Department of Mathematical Sciences, Durham University, Upper Mountjoy, Stockton Road, Durham DH1 3LE, UK}
\affiliation{ Faculty of Sciences, Hasselt University, Agoralaan Gebouw D, Diepenbeek, 3590 Belgium}

\begin{abstract}
Timing and phase resolution in satellite QKD, kilometre-scale gravitational-wave detectors, and space-borne clock networks hinge on quantum-speed limits (QSLs), yet benchmarks omit relativistic effects for coherent and squeezed probes. We derive first-order relativistic corrections to the Mandelstam-Tamm and Margolus-Levitin bounds. Starting from the Foldy-Wouthuysen expansion and treating $-p^{4}/(8 m^{3} c^{2})$ as a harmonic-oscillator perturbation, we propagate Gaussian states to obtain closed-form QSLs and the quantum Cram\'er-Rao bound. Relativistic kinematics slow evolution in an amplitude- and squeezing-dependent way, increase both bounds, and introduce an $\epsilon^{2} t^{2}$ phase drift that weakens timing sensitivity while modestly increasing the squeeze factor. A single electron ($\epsilon \approx 1.5\times 10^{-10}$) in a $5.4\,\mathrm{T}$ Penning trap, read out with $149\,\mathrm{GHz}$ quantum-limited balanced homodyne, should reveal this drift within $\sim 15\,\mathrm{min}$ -- within known hold times. These results benchmark relativistic corrections in continuous-variable systems and point to an accessible test of the quantum speed limit in high-velocity or strong-field regimes.
\end{abstract}

\maketitle

\section*{Introduction}

How fast can quantum information or optical phase accumulate when the
carriers themselves approach relativistic speeds?  
That question defines the {fundamental} limits on timing and phase
resolution. These limits are central to satellite quantum-key-
distribution links \cite{Liao2017SatQKD}, laser-interferometric
gravitational-wave detectors \cite{Abbott2016GW}, and emerging
space-borne clock networks \cite{Komar2014}.  
The key yardstick is the {quantum-speed limit} (QSL): the minimum
time for a quantum state to evolve into a distinguishable form.
Mandelstam and Tamm first derived it \cite{Mandelstam1945}, and
Margolus and Levitin later generalised it \cite{Margolus1998}.  
Here we apply the leading relativistic correction,
\( -p^{4}/(8m^{3}c^{2}) \) from the Foldy-Wouthuysen expansion, and
re-evaluate the QSLs for the two Gaussian workhorses of quantum
optics-coherent and squeezed states.  
Non-relativistic QSLs are well-catalogued \cite{Deffner2017Review}, but
a clear formula is still missing for how the quartic term alters the
minimum gate time or phase sensitivity.  
We close that gap and outline a balanced-homodyne protocol capable of
experimentally testing the predicted corrections with present-day
hardware.

\smallskip
Robertson’s generalised uncertainty relation sets an upper limit on how
fast, the expectation value of any observable can change
\cite{Robertson1929}.  Applied to the time-energy pair, the relation
yields the Mandelstam-Tamm lower bound on a closed system’s evolution
time \cite{Mandelstam1945}; Margolus and Levitin later introduced a
complementary bound that depends on the mean energy above the ground
state \cite{Margolus1998}.  Together, these two results define the
modern {quantum-speed limits} and have since been extended to
mixed, correlated, and non-Markovian dynamics
\cite{Deffner2013,Taddei2013,Smerzi2012}.  Geometric approaches link
the same quantities to the statistical distance on Hilbert space and to the
Quantum Fisher information that fixes metrological precision
\cite{Anandan1990,Braunstein1994,Florian2015}.  
We use these non-relativistic foundations as the reference point for
our forthcoming relativistic analysis\cite{Aznabayev2019} of coherent and squeezed states.

\smallskip
Beyond their foundational interest, quantum-speed limits are also
crucial for metrology: the same time-energy trade-off sets the quantum
Fisher information (QFI), which in turn determines the ultimate
precision given by the quantum Cramér-Rao bound
\cite{Florian2015,Giovannetti2011}.  
Kilometre-scale gravitational-wave interferometers illustrate this
point: injecting a squeezed vacuum raises their strain sensitivity
beyond the standard quantum limit
\cite{Aasi2013Enhanced,Acernese2019Squeezed}.  The QSL-defined
interrogation time, then limits how quickly phase information can be
gathered and read out.  
Similar timing constraints apply to satellite
quantum-key-distribution links and space-borne clock networks, where
photonic qubits follow relativistic trajectories
\cite{rideout2012fundamental,Liao2017SatQKD}.  
Optical-clock networks that search for ultralight-scalar dark matter
face the same restriction: even when noise is modelled with standard
shot-noise formulas, the smallest dilaton-induced frequency shift that
can ever be detected is still bounded by the QFI-and therefore by QSLs
\cite{Arvanitaki2015}.

\smallskip
Despite their broad reach, nearly all analytic studies assume
non-relativistic dynamics.  Only two exceptions exist: a bound for a
single Dirac electron in a uniform magnetic field
\cite{Villamizar2015QuantumSpeed}, and heuristic estimates based on
generalised-uncertainty relations at the Planck scale
\cite{Bosso2017GUP}.  No closed-form relativistic limit has yet been
derived for the Gaussian mainstays of quantum optics-coherent and
squeezed states-even though they underpin satellite timing protocols
\cite{rideout2012fundamental} and supply the squeezed vacuum that
enhances ground-based interferometers \cite{Aasi2013Enhanced}.
Consequently, the community still lacks a benchmark that shows how the
leading relativistic term  alters both the
minimum evolution time and the best achievable phase sensitivity.

\smallskip
Here we derive closed-form, first-order relativistic corrections to the
Mandelstam-Tamm and Margolus-Levitin quantum-speed limits for coherent
and squeezed light using the \( p^{4}/(8m^{3}c^{2}) \) term from the
Foldy-Wouthuysen expansion \cite{Foldy1950}.  These results yield the
corresponding relativistic quantum Cramér-Rao bound, reveals how the
correction modifies the squeeze factor, and identify a balanced-homodyne
experiment that can detect the predicted \( \epsilon^{2}t^{2} \) timing
drift with existing hardware. {Related metrological analyses also exist within {deformed-commutator} (GUP/MCR) frameworks,
where minimal-length or maximum-momentum effects are encoded in a modified $[X,P]$ and quantum
probes are used to estimate the {deformation parameter} via QFI~\cite{ChingNg2013PRD,
RossiGianiParis2016PRD,ChingNg2019PRD,CandeloroParis2020PRD}. In contrast, here we retain the
{canonical} algebra and add only the Foldy-Wouthuysen relativistic kinetic correction,}

\smallskip
We start from the relativistic harmonic-oscillator Hamiltonian that
emerges from the Foldy-Wouthuysen expansion of the Dirac equation.  Its
leading correction adds the quartic term
 to the standard oscillator
\cite{Foldy1950,Dirac1928}.  Using perturbation theory
\cite{Messiah1962}, we obtain corrected eigenvalues and ladder
operators, propagate coherent and squeezed states, and extract the
Mandelstam-Tamm and Margolus-Levitin bounds from their Fubini-Study
distance \cite{Anandan1990,Jones2010GeometricDerivation}.  The same
calculations provide the energy moments needed for the relativistic
quantum Fisher information, which leads directly to the modified
Cramér-Rao bound and its effect on optical squeezing.  To bridge theory
and experiment, we analyse the phase sensitivity of balanced homodyne
detection.  This method can saturate the Cramér-Rao limit
\cite{YuenChan1983} and has already been realised with a quantum-limited
149\,GHz SIS mixer \cite{Shan2019BalancedSIS}.

\smallskip
Our results reveal three key trends.  
\textit{(i)} For {coherent states}, relativistic kinematics
shorten the intervals of accelerated evolution but beyond a threshold
amplitude, enforce an overall slowdown; hence, the fastest gate time for
photon-number qubits increases.  
\textit{(ii)} For {squeezed states}, the same
 relativistic correction term raises both the Mandelstam-Tamm and
Margolus-Levitin times while slightly increasing the squeeze factor,
providing extra quantum-noise suppression precisely where timing becomes
stricter.  
\textit{(iii)} Propagating these corrections to metrology yields a
relativistic quantum Cramér-Rao bound whose reduced sensitivity appears
as an \( \epsilon^{2}t^{2} \) phase drift in balanced-homodyne read-out.
Numerical estimates indicate that this drift rises above shot noise
after roughly 15\,min in a 5.4\,T Penning trap equipped with a
quantum-limited 149\,GHz SIS mixer
\cite{Hanneke2008,Shan2019BalancedSIS}, placing the effect within reach
of current technology. { We integrate the phase effect into a standard Gaussian-modulated coherent-state (GMCS) CV-QKD model with a locally generated local oscillator (LLO) and pilots, derive the input-referred excess-noise term, and quantify the resulting key-rate penalty within the standard security framework \cite{Weedbrook2012RMP}.
This connects the relativistic drift to the dominant phase-noise budget in satellite CV-QKD and suggests drop-in mitigations--raise pilot SNR, shorten the estimation window, and apply linear prediction--compatible with current LLO architectures \cite{Dequal2021,Kish2020QE,Shao2021PRA,Laudenbach2019Quantum,Marie2017PRA}.}

\section{Relativistic Corrections}
Now we review the first-order relativistic corrections to a particle in a harmonic oscillator. 
The relativistic Hamiltonian for a particle in a harmonic oscillator potential combines relativistic kinetic energy with the harmonic potential energy,  expressed as $H_{\text{rel}} = \sqrt{p^2 c^2 + m^2 c^4} - mc^2 + \frac{1}{2} m \omega^2 x^2$   \cite{Dirac1928}, where $p$ represents the momentum operator, $m$ is the mass, $c$ is the speed of light, $\omega$ is the angular frequency, and $x$ is the position operator. The term $\sqrt{p^2 c^2 + m^2 c^4}$ captures the relativistic kinetic energy with the rest energy subtracted, while $\frac{1}{2} m \omega^2 x^2$ is the harmonic potential.
For low-momentum, where relativistic effects are small, the Hamiltonian can be expanded to the first order in $\frac{p^2}{m^2 c^2}$. This yields $H_{\text{rel}} \approx \frac{p^2}{2m} + \frac{1}{2} m \omega^2 x^2 - \frac{p^4}{8 m^3 c^2}$   \cite{Bjorken1964}. Here, the first two terms correspond to the standard non-relativistic harmonic oscillator Hamiltonian, while the term $-\frac{p^4}{8 m^3 c^2}$ represents the leading-order relativistic correction. This correction becomes significant when the particle's momentum approaches relativistic values.
 We can write the relativistically corrected Hamiltonian as $H = H_0 - \epsilon \delta H$, where $H_0 = \frac{p^2}{2} + \frac{x^2}{2}$ is the non-relativistic Hamiltonian, $\epsilon = \frac{1}{m^3 c^2}$ is a small parameter quantifying relativistic effects, and encapsulates the relativistic perturbation. This decomposition allows the corrections to be treated systematically within the framework of perturbation theory   \cite{Messiah1962}.
The energy eigenvalues of the unperturbed harmonic oscillator are given by $E_n^{(0)} = n + \frac{1}{2}$   \cite{Landau1977}, where $n$ is the quantum number. These equally spaced levels reflect the hallmark property of the harmonic oscillator. Relativistic corrections are incorporated by evaluating the expectation value of $\delta H$ in the unperturbed eigenstates. Taking the first-order perturbation into consideration, the total energy is given by $E_n=E_n^{(0)} -\epsilon E_n^{(1)} $, where $E_n^{(1)} = \frac{1}{32}(6n^2 + 6n + 3)$ (see Appendix A). The quadratic dependence of the correction on $n$ reflects the increasing importance of relativistic effects as the system's energy increases.

The fundamental operators of the oscillator, including the annihilation and creation operators, are also modified to reflect relativistic effects. Up to the first order in $\epsilon$, the annihilation operator becomes $a = a_0 + \frac{\epsilon}{32} \left( -2 a_0^3 + 6 N_0 a_0^\dagger - (a_0^\dagger)^3 \right)$   \cite{Dirac1928}, where $a_0$ and $a_0^\dagger$ are the unperturbed operators, and $N_0 = a_0^\dagger a_0$ is the number operator. Therefore, the modified position and momentum operators can be expressed as (see Appendix A)  $x = \frac{1}{\sqrt{2}} \left( a_0^\dagger + a_0 \right) + \frac{3 \epsilon}{32 \sqrt{2}} \left( a_0^3 + (a_0^\dagger)^3 - 2 a_0 N_0 - 2 N_0 a_0^\dagger \right)$,  $p = \frac{i}{\sqrt{2}} \left( a_0^\dagger - a_0 \right) - \frac{3 i \epsilon}{32 \sqrt{2}} \left( a_0^3 - (a_0^\dagger)^3 \right)$. These corrections demonstrate how relativistic effects perturb the canonical operators, leading to changes in the oscillator's dynamics and spectrum. 
Thus, the modified Hamiltonian can be expressed in terms of the modified number operator, $N\equiv a^\dagger a$, as $H = N + \frac{1}{2} - \frac{\epsilon}{32} \left( 6 N^2 + 6 N + 3 \right)$. The additional quadratic dependence on $N$ introduces a nonlinear relationship between the relativistically corrected quantum number and energy levels, pointing towards the growing influence of relativistic effects with increasing $n$ (see Appendix A). 

This behavior is consistent with the relativistic corrections observed in more complex systems   \cite{Bjorken1964}. These results demonstrate how relativistic dynamics are manifested even in seemingly simple quantum systems such as the harmonic oscillator. We use these results to study such corrections to QSLs for coherent and squeezed states considering their importance and significance.

To propagate a displaced oscillator we re-express the initial Glauber
state in the perturbed number basis
\(\{\ket{n}\}\).  Retaining the usual Poissonian weights,
the time-evolved wave-packet reads  \cite{Bosso2017GUP}
\begin{equation}
\label{eq:alpha_prime}
\ket{\alpha'(t)}
  = \exp\!\bigl[-|\alpha(0)|^{2}/2\bigr]
    \sum_{n=0}^{\infty}
      \frac{\alpha(0)^{\,n}}{\sqrt{n!}}\;
      e^{-iE_{n}t}\ket{n},
\end{equation}
with {\(E_{n}=E_{n}^{(0)}-\epsilon E^{(1)}_{n}\)} the corrected spectrum (from
Appendix~B).  Equation~\eqref{eq:alpha_prime} shows that relativistic
effects enter exclusively through (i) the modified eigenfunctions
\(\ket{n}\) and (ii) the phase factors
\(\exp[-iE_{n}t]\); the photon-number distribution itself is
unchanged. Applying the corrected squeeze operator \(s(\varsigma)\) to the
relativistic vacuum \(\ket{0}\) gives  \cite{Bosso2017GUP}
\begin{equation}
\label{eq:sq_prime}
\begin{aligned}
\ket{\varsigma(t)} &=
  \sum_{n=0}^{\infty}
    f(2n)\,e^{-iE_{2n}t}\ket{2n},\\[2pt]
f(2n) &=
  \frac{\sqrt{(2n)!}\,(-\tanh r)^{n}}
       {2^{n} n!\sqrt{\cosh r}}.
\end{aligned}
\end{equation}
where \(E_{2n}=E_{2n}^{(0)}-\epsilon E^{(1)}_{2n}\).
The coefficients \(f(2n)\) coincide with the non-relativistic case, so
the quadrature variances-and hence the squeezing advantage-are fully
preserved, while the relativistic correction manifests solely as a
phase shift in each even-photon component.
\section*{ Quantum Speed Limits with First Order Relativistic Corrections}
Quantum speed limits (QSLs) impose fundamental bounds on the evolution time required for a quantum system to evolve between two distinguishable states. In this work, we consider two widely studied speed limits bounds for non-relativistic unitary dynamics, namely Mandelstam-Tamm (MT) bound~  \cite{Mandelstam1945} and Margolus-Levitin (ML) bound  \cite{Margolus1998}. The unified bound for arbitrary initial and final states connected via a unitary transformation is given by 
\begin{equation}
 T_{\text{QSL}} = \max\{T_{\text{MT}},\, T_{\text{ML}}\}
\end{equation}
where $T_{\text{MT}} = {\arccos\left(|\langle \psi(0)|\psi(t)\rangle|\right)}/{\Delta E}$  with \(\Delta E\) is the uncertainty in energy of the state and
{  $ T_{\text{ML}} = {2\arccos\left(|\langle \psi(0)|\psi(t)\rangle|\right)^2}/{2\langle E \rangle \pi}$}.

We now incorporate the relativistic corrections to the ML and MT bounds. Since in this work, we are dealing with continuous variable systems, we consider the two different types of states of such systems, namely coherent states~  \cite{Zhang1990Coherent} and squeezed states,~  for our work due to their usefulness in various context as discussed above. We first consider the coherent states, and the introduction of relativistic corrections alters the energy spectrum and thus the speed of evolution at which a system can evolve. The ML bound with relativistic correction for coherent states  can be expressed as (see Appendix B):
\begin{align}
T_{\text{ML}} &= T_{\text{ML0}}
               + \frac{3\,\epsilon\,w_1}{4\,w_3\,\pi}
                 \left(
                   w_4\,w_1 - \frac{w_5\,w_6\,w_7\,w_8\,w_9}{w_{10}}
                 \right)
\end{align}
$ T_{\text{ML0}} =\frac{2\,w_1^{2}}{w_2\,\pi},
w_1 = \arccos\!\bigl(e^{\alpha_0^{2}\,(-1+\cos t)}\bigr), 
w_2 = \tfrac12 + \alpha_0^{2},
w_3 = \bigl(1 + 2\alpha_0^{2}\bigr)^{2}, 
w_4 = 1 + 4\alpha_0^{2} + 2\alpha_0^{4},
w_5 = 4\alpha_0^{2}\bigl(1 + 2\alpha_0^{2}\bigr), 
w_6 = e^{\alpha_0^{2}\,(-1+\cos t)},
w_7 = t,
w_8 = 1 + \alpha_0^{2}\cos t,
w_9 = \sin t, 
w_{10} = \sqrt{1 - e^{2\alpha_0^{2}\,(-1+\cos t)}}. $ Here \(\alpha_0 = |\alpha(0)|\) denotes the magnitude of the initial coherent-state parameter, $  T_{\text{ML0}}$ is the non-relativistic limit, and \(\epsilon\) is the small relativistic correction parameter.
The first term corresponds to the standard ML bound, whereas the second term introduces higher-order relativistic corrections. These corrections highlight the decrease in time-sensitivity. The presence of $ \epsilon $ alters the dependence of the ML bound on both the coherent state amplitude and the evolution time.
Similarly, the MT bound with relativistic correction for coherent states can be expressed as  (see Appendix B):
\begin{align}
T_{\text{MT}} &= T_{\text{MT0}}
               + \frac{3\,\epsilon}{8}
                 \left(
                   \frac{v_3\,v_1}{v_2}
                   - \frac{v_2\,v_4\,v_5\,v_6\,v_7}{v_8}
                 \right),
\end{align}

$ T_{\text{MT0}} =\frac{v_1}{v_2},
v_1 = \arccos\!\bigl(e^{\alpha_0^{2}\,(-1+\cos t)}\bigr), 
v_2 = \alpha_0,
v_3 = 1+\alpha_0^{2}, 
v_4 = e^{\alpha_0^{2},(-1+\cos t)},
v_5 = t,
v_6 = 1+\alpha_0^{2}\cos t,
v_7 = \sin t,
v_8 = \sqrt{1 - e^{2\alpha_0^{2}\,(-1+\cos t)}}.
$ Here \(T_{\text{MT0}}\) is the Mandelstam-Tamm bound in the non-relativistic limit, while the bracketed term multiplied by \(\epsilon\) gives the first-order relativistic correction.

In this expression, the first term corresponds to the traditional MT bound, which scales inversely with the uncertainty in energy. The second term introduces relativistic corrections proportional to $ \epsilon $, which depend on both the initial amplitude $ \alpha_0 $ and the evolution time $ t $. These corrections also reduce the minimum evolution time for the system.
As QSL is the maximum of the ML and MT bounds, $T_{QSL}= \text{max}\{T_{\text{ML}},T_{\text{MT}}\}$, we can say that due to relativistic effects, the minimum time of evolution between the coherent states increases.

We can now also analyze the QSLs of squeezed states   \cite{Schnabel2017Squeezed}. The corrected ML bound for squeezed can be expressed as (see Appendix C): 
\begin{align}
    T_{\text{ML}} &= 
        T_{\text{ML0}} 
    + \frac{3\epsilon\, x_3}{16 \pi} x_1 \left[ x_4\, x_3\, x_1 
    + x_5\, \frac{x_6}{x_7^{1/4}\, x_8}\right],
\end{align} 
$ T_{\text{ML0}}=\frac{4\,x_3}{\pi} x_1^2 $ is the standard limit, and  the  terms \(x_i\) are 
$    x_1 = \arccos\left(\frac{\sqrt{2}}{x_2^{1/4}}\right), 
    x_2 = 3 + \cosh(4r) - 2 \cos t\,\sinh^2(2r), 
    x_3 = \text{sech}(2r), 
    x_4 = 1 + 3\cosh(4r), 
    x_5 = \frac{32\,t\,\cosh^5 r\,\sinh^2 r}{x_2^{7/4}}, 
    x_6 = -4\sin t + \sin(2t)\,\tanh^2 r + 2\sin t\,\tanh^4 r, 
    x_7 = 1 - 2\cos t\,\tanh^2 r + \tanh^4 r, 
    x_8 = \sqrt{-2 + \sqrt{x_2}}.
$
This expression reveals a dependence of the ML bound on the squeezing parameter $r$ and the relativistic correction parameter $\epsilon$. The presence of hyperbolic functions, such as $\cosh(4r)$ and $\sinh^2(2r)$, which arise from the squeezing process, amplifies the impact of relativistic effects. These corrections result in tighter constraints on the evolution time, particularly for highly squeezed states where $r$ is large. Physically, this implies that systems with high degrees of squeezing are more sensitive to relativistic effects, which can either enhance or limit their utility in high-precision applications.
The MT bound, which is governed by the uncertainty in energy, also experiences significant modifications due to relativistic corrections. The corrected MT bound for the squeezed state is  given by (see Appendix C): 
\begin{align}
    T_{\text{MT}} &= T_{\text{MT0}}+ \frac{3 \epsilon\,y_3}{16\sqrt{2}} 
    \left[\,6\,y_4\,y_1 + y_5\,\frac{y_6}{y_7^{1/4}\,y_8}\right],
\end{align}
  With $ T_{\text{MT0}}=y_3\,y_1 $  the unperturbed MT bound, and  the defined terms \(y_i\) are:
$
    y_1 = \arccos\left(\frac{\sqrt{2}}{y_2^{1/4}}\right), 
    y_2 = 3 + \cosh(4r) - 2 \cos t\,\sinh^2(2r), 
    y_3 = \sqrt{2}\,\text{csch}(2r), 
    y_4 = \cosh(2r), 
    y_5 = \frac{8\,t\,\cosh^5 r\,\sinh^2 r}{y_2^{7/4}}, 
    y_6 = -4\sin t + \sin(2t)\,\tanh^2 r + 2\sin t\,\tanh^4 r, 
    y_7 = 1 - 2\cos t\,\tanh^2 r + \tanh^4 r, 
    y_8 = \sqrt{-2 + \sqrt{y_2}}.
$
The corrected MT bound demonstrates how the interaction between energy uncertainty, squeezing, and relativistic corrections affects the quantum evolution time. The dependence on $\epsilon$ and $r$ indicates that highly squeezed states, while beneficial for improving measurement precision, are more susceptible to relativistic effects. This sensitivity can impose additional constraints on their evolution and utility in relativistic quantum systems.

\begin{figure}[H]  \centering
  \includegraphics[width=\columnwidth]{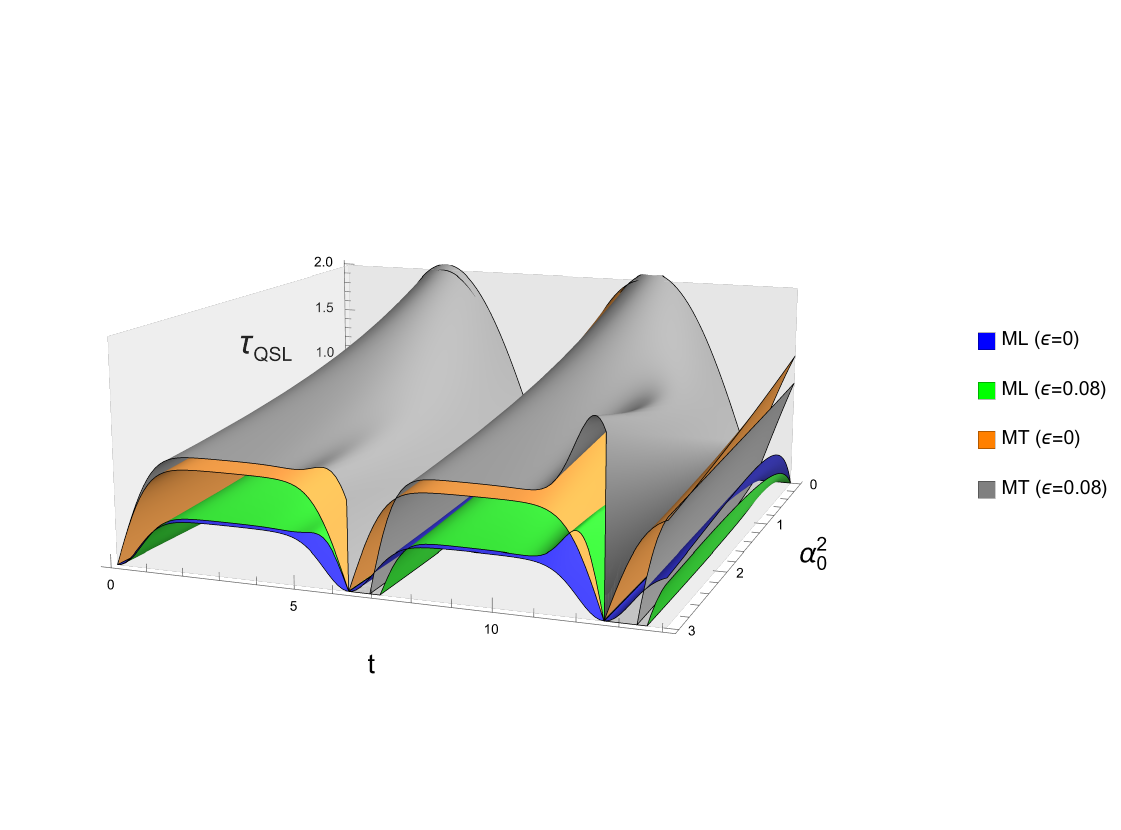}
  \caption{\textbf{First-order relativistic modification of MT and ML bounds for coherent states.}
Surfaces give the Mandelstam-Tamm (MT) and Margolus-Levitin (ML) bounds
as functions of the evolution time \(t\) and the mean photon number
\(\alpha_{0}^{2}\).  Blue and orange sheets are the non-relativistic
limits (\(\epsilon=0\)); green and grey include the first-order
relativistic correction \(\epsilon=0.08\).
Throughout the plotted window 
\(\mathrm{MT}^{(\epsilon)}\) (grey) forms the
true quantum speed limit.}
  \label{fig:plot1}
\end{figure}

{For coherent states, Fig.~\ref{fig:plot1} plots the Margolus-Levitin and Mandelstam-Tamm speed-limit times versus the initial amplitude \(\alpha_{0}\) and time \(t\). The first-order relativistic correction grows with \(\alpha_{0}\) and oscillates with \(t\), changing sign only within narrow phase windows. As \(\alpha_{0}\) increases, these windows shrink, so the overall effect is a slow-down (the bounds increase). A narrow pre-drop spike appears just before each revival at \(t\simeq 2\pi n\). This is a near-revival artifact \cite{OlverEtAl2010}: as the fidelity approaches unity, \(F(t)=\exp[-\alpha_{0}^{2}(1-\cos t)]\to 1\), the Bures/Fubini-Study angle \(\Theta(t)=\arccos F(t)\) shows square-root scaling, \(\Theta(t)\simeq \sqrt{2\,[1-F(t)]}\simeq \alpha_{0}\,|t-2\pi n|\). The result is a sharp, non-resonant enhancement whose height increases with \(\alpha_{0}\).}

\begin{figure}[H]
  \centering
  \includegraphics[width=\columnwidth]{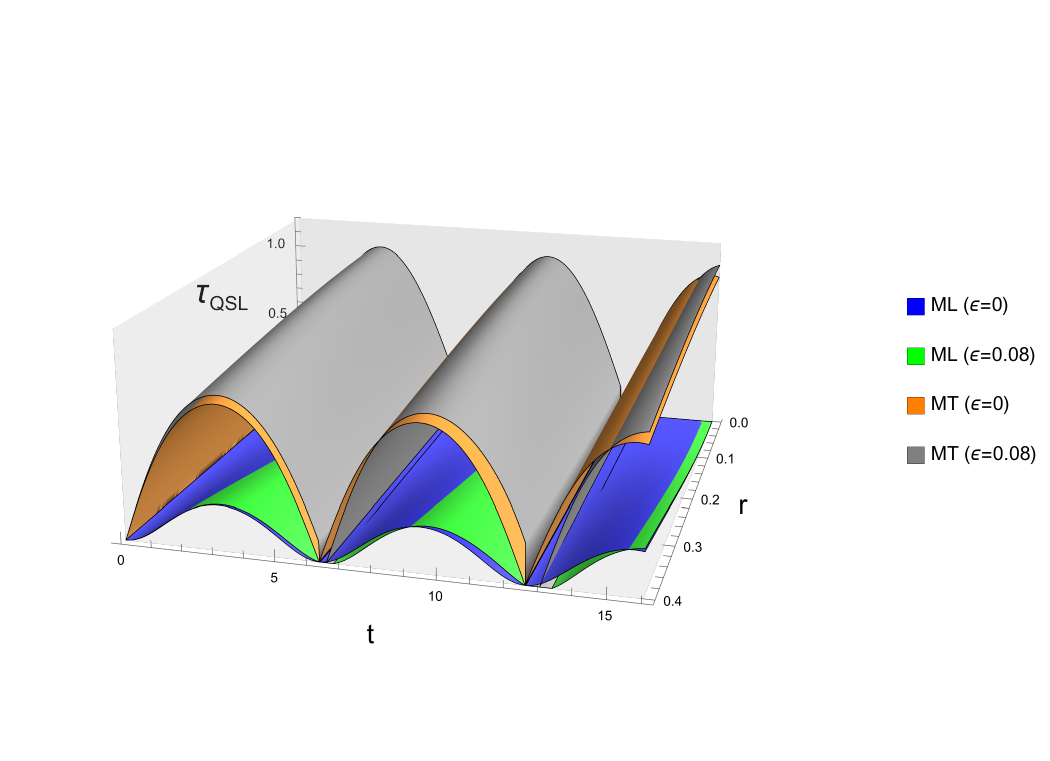}
  \caption{\textbf{First-order relativistic modification of MT and ML bounds for squeezed vacua.}
Same conventions as Fig.~\ref{fig:plot1}, but with the squeezing
parameter \(r\) on the secondary axis \((0\le r\le0.4)\).
First-order relativistic corrections (\(\epsilon=0.08\))
lift both ML (green) and MT (grey) bounds over the entire domain,
and the gap between corrected and uncorrected surfaces widens
monotonically with \(r\), signalling the enhanced sensitivity of
highly squeezed states to relativistic effects.}
  \label{fig:plot}
\end{figure}

{For squeezed inputs, the relativistic correction increases both the Margolus-Levitin (ML) and Mandelstam-Tamm (MT) bounds, slowing the evolution. On average, the increase grows with the squeezing parameter \(r\). Brief intervals persist where the correction is negative (a transient speed-up), but their duration decreases as \(r\) increases. This trend matches the plots: the corrected ML/MT surfaces lie mostly above the unperturbed ones, with the remaining sign changes tied to the phase of the unperturbed oscillations (see Fig.~\ref{fig:plot}).}

 Squeezed states redistribute quantum noise below the shot‐noise level, making them indispensable for enhancing the strain sensitivity of ground-based interferometers such as LIGO and Virgo   \cite{Abbott2016} and for boosting signal-to-noise in quantum‐enhanced spectroscopy and imaging   \cite{Caves1981}.  Relativistic corrections modify the minimal evolution times prescribed by the QSLs, altering how quickly squeezed light can accumulate or read out phase information in high-velocity platforms (e.g.\ satellite gravitational-wave detectors) or strong‐field environments (e.g.\ near black-hole binaries).  These corrections also increase the squeeze factor by slightly amplifying non-classical noise suppression, thereby extending the reach of squeezed-state sensors while simultaneously dictating the fastest operations they can perform in extreme relativistic scenarios.

\section{Quantum Cramér-Rao Bound with Relativistic Correction}
We now analyze the relativistic corrections to the Quantum Cramér-Rao bound   \cite{braunstein1994statistical}.  The Quantum Cramér-Rao bound sets a fundamental limit on how precisely one can estimate a parameter that is encoded in a quantum state. It states that for any unbiased estimator of a parameter, the variance (or uncertainty) of the estimation has a lower bound equivalent to the inverse of the quantum Fisher information (QFI). This bound incorporates the effects of the quantum nature of the system-specifically, the non-commuting observables and the probabilistic outcomes of quantum measurements-and therefore provides a limit on the best possible precision achievable in quantum metrology and quantum parameter estimation. Relativistic corrections are important to investigate this bound under high-speed or even strong gravitational conditions, thereby aligning quantum metrology with relativistic physics. For relativistically corrected coherent states, the energy variance is modified to 
$
\Delta H_{\ket{\alpha'}}^2 = \alpha_0^2 - \frac{3}{4}\left(\alpha_0^2 + \alpha_0^4\right)\epsilon$. This reduction in variance lowers the Quantum Fisher Information (QFI) to $ \mathcal{F}_t = 4\alpha_0^2 - 3\left(\alpha_0^2 + \alpha_0^4\right)\epsilon$, thereby increasing the Quantum Cramér-Rao bound (QCRB) for time parameter estimation, i.e., $ \mathrm{QCRB} \geq \frac{1}{\sqrt{\mathcal{F}_t}}$.

\begin{figure}[!t]
  \captionsetup{position=top}
  \captionsetup[subfigure]{justification=centering}
  \centering

  \begin{subfigure}[b]{0.48\columnwidth}
    \includegraphics[width=\linewidth]{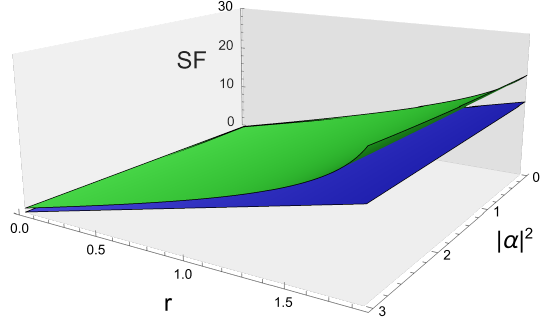}
    \caption{$\theta = 0$}
    \label{fig:sub1}
  \end{subfigure}
  \hfill
  \begin{subfigure}[b]{0.48\columnwidth}
    \includegraphics[width=\linewidth]{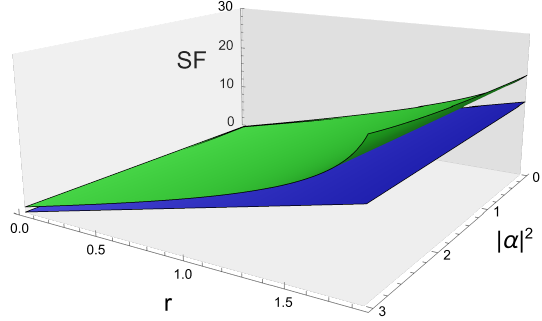}
    \caption{$\theta = \pi/4$}
    \label{fig:sub2}
  \end{subfigure}

  \vspace{1em}

  \begin{subfigure}[b]{0.48\columnwidth}
    \includegraphics[width=\linewidth]{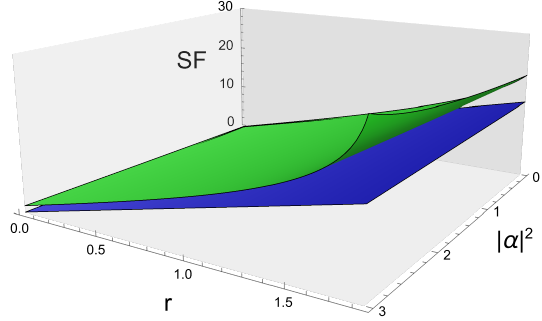}
    \caption{$\theta = \pi/2$}
    \label{fig:sub3}
  \end{subfigure}
  \hfill
  \begin{subfigure}[b]{0.48\columnwidth}
    \includegraphics[width=\linewidth]{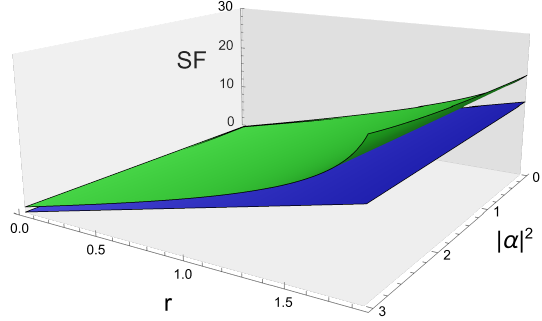}
    \caption{$\theta = 3\pi/4$}
    \label{fig:sub4}
  \end{subfigure}

  \caption{\textbf{Relativistic enhancement of the optical squeeze factor.}
  Squeeze-factor (SF) surfaces, defined by
  \(\mathrm{SF}=-10\log_{10}\bigl[\Delta^{2}X_{s}/\Delta^{2}X_{c}\bigr]\),
  are plotted versus \(|\alpha|^{2}\) and the squeezing parameter \(r\)
  for four relative phases \(\theta\) between displacement and squeezing.
  Blue sheets: non-relativistic (\(\epsilon=0\)); green sheets:
  first-order relativistic correction (\(\epsilon=0.08\)).
  Relativity lifts the SF across the full domain
  (\(0\le r\le1.8,\;0\le|\alpha|^{2}\le3\)),
  with a phase-dependent magnitude.}
  \label{fig:main}
\end{figure}

As regimes become more relativistic or the amplitude of coherent states increases, so does the impact on the sensitivity of quantum sensors and metrological devices. Similarly, for the squeezed states, the energy variance is given by
$
\Delta H_{\ket{\varsigma}}^2 = 2\cosh^2(r)\sinh^2(r) - \frac{9}{32}\sinh(2r)\sinh(4r)\epsilon,
$
with \(r\) as the squeezing parameter. Thus, the QFI becomes
$
\mathcal{F}_t = 8\cosh^2(r)\sinh^2(r) - \frac{9}{8}\sinh(2r)\sinh(4r)\epsilon.
$ The relativistic correction terms, scaling with \(r\), lead to a less tight QCRB, indicating reduced time sensitivity.
On the other hand, the Heisenberg uncertainty principle for non-relativistic minimum uncertainty  states is expressed as
$
\Delta^2 X \Delta^2 P \ge \frac{1}{16},
$
with the position quadrature variance fixed at \(\Delta^2 X = \frac{1}{4}\). In squeezed states, where \(\Delta^2 X < \frac{1}{4}\), the degree of nonclassicality is quantified by the squeeze factor,
$
    -10 \cdot \log_{10}\left(\frac{\Delta^2 X_s}{\Delta^2 X_c}\right),
$
with \(e^{-2r} = \frac{\Delta^2 X_s}{\Delta^2 X_c}\) serving as the squeezing parameter. Incorporating relativistic corrections, we obtain
\begin{equation}
    \frac{\Delta^2 X_s}{\Delta^2 X_c} = e^{-2r} - \frac{3}{64}\epsilon \left( 5 + 3e^{-4r} - 4\alpha_0^2 e^{-2r} (\cos(2\theta)-4) \right),
\end{equation}
where \(\Delta^2 X_c = \frac{1}{4}\) is the non-relativistic benchmark. As illustrated in Fig. \ref{fig:main}, relativistic effects cause a slight increase in the squeeze factor, which leads to enhanced noise suppression and optical gain. {Relativistic corrections reduce time sensitivity while modestly enhancing the state’s nonclassical squeezing (larger SF). The effect arises from unitary covariance reshaping under the FW term and does not presume any external noise}

\FloatBarrier

\section{Balanced-homodyne measurement of the relativistic phase drift}
\begin{figure*}
    \centering
    \includegraphics[width=0.8\linewidth]{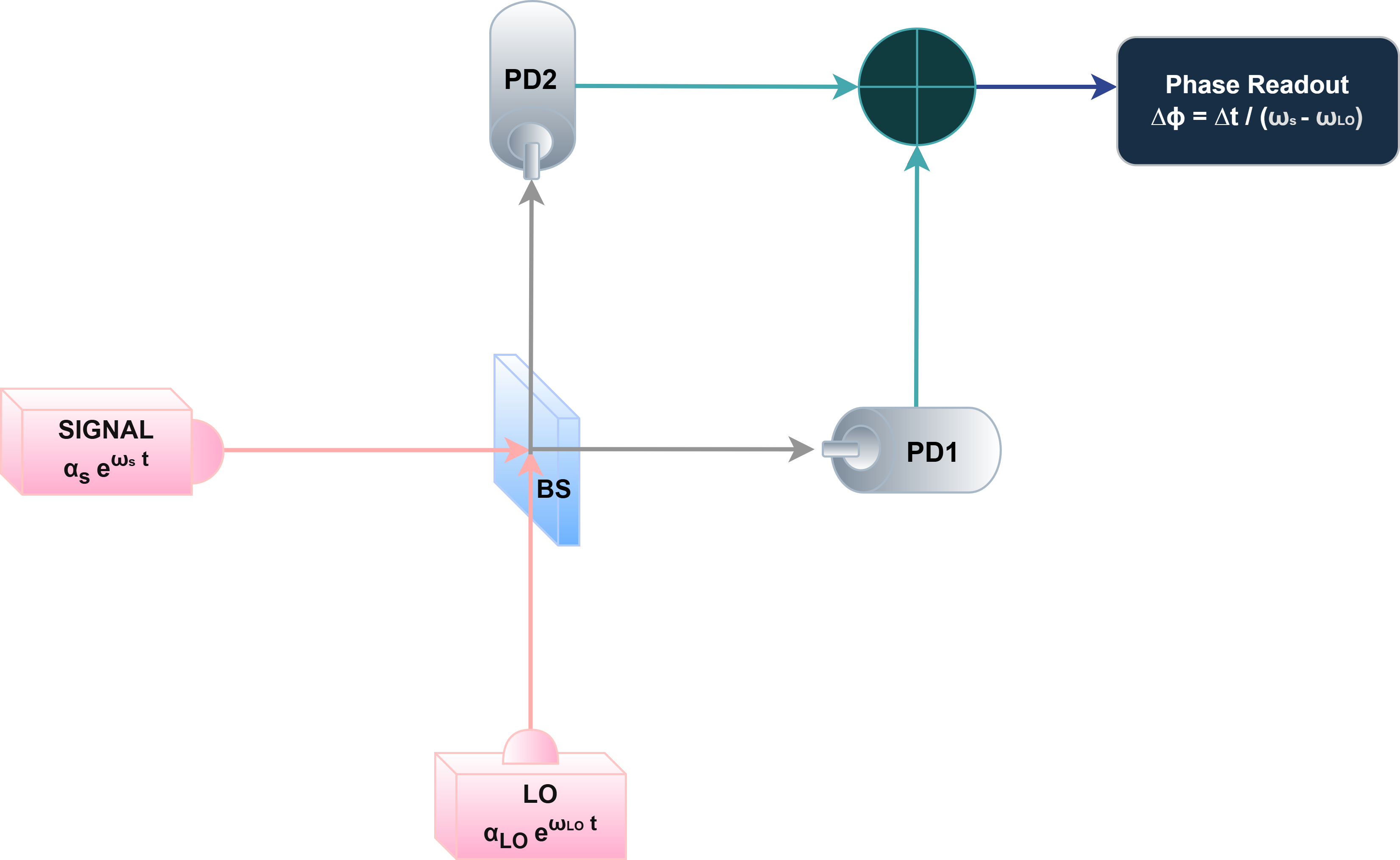}
    \caption{BHD scheme for experimentally testing the relativistic corrections to QSLs.}
    \label{fig:enter-label}
\end{figure*}

In this section, we propose a balanced homodyne detection experiment to measure the relativistic drift in phase sensitivity (and the corresponding time-resolution limit), thereby indirectly probing the predicted corrections to the quantum speed limit bounds. BHD is well suited to detect such corrections, as it is a widely used technique in quantum optics for measuring the phase of optical fields with high precision   \cite{bhd, bhd1}. In this approach, a signal field is mixed with a strong Local Oscillator (LO) field at a $50:50$ beam splitter. The difference in intensities at the output ports of the beam splitter is measured, allowing the extraction of the phase difference between the signal and LO fields. This is illustrated in  Fig. \ref{fig:enter-label}. The phases of the time-evolving signal and LO are given by $ \psi_\text{sig}(t) = \omega_s t $ and $ \psi_\text{LO}(t) = \omega_\text{LO} t $, respectively. The relative phase difference between the signal and the LO is defined as 
 $
    \delta \psi = \psi_\text{sig}(t) - \psi_\text{LO}(t) = (\omega_s - \omega_\text{LO}) t.
 $

The system consists of two modes: the Signal Mode and the Local Oscillator (LO) Mode. The Signal Mode is described by the annihilation operator $\hat{a}_s$ and is prepared in a coherent state $|\alpha_s e^{i \psi_s}\rangle$. The LO Mode is defined by a relativistically corrected annihilation operator $ \hat{a}_{\text{LO}} $ and is in a modified coherent state $|\alpha_{\text{LO}} e^{i \psi_{\text{LO}}}\rangle$. The modified annihilation operator is expressed as 
$ \hat{a}_{\text{LO}} = a_0 + \frac{\epsilon}{32} \left( -2 a_0^3 + 6 N_0 a_0^\dagger - (a_0^\dagger)^3 \right) $, 
where $\epsilon$ represents the relativistic correction parameter. These corrections account for relativistic effects, which become significant in high-energy regimes.

At the beam splitter, the input fields are combined using the relations $ \hat{a}_{\text{out1}} = \frac{\hat{a}_s + \hat{a}_{\text{LO}}}{\sqrt{2}} $ and $ \hat{a}_{\text{out2}} = \frac{\hat{a}_s - \hat{a}_{\text{LO}}}{\sqrt{2}} $. The difference intensity operator is defined as 
$ \hat{I}_{\text{diff}} = \hat{a}_{\text{out1}}^\dagger \hat{a}_{\text{out1}} - \hat{a}_{\text{out2}}^\dagger \hat{a}_{\text{out2}} $, 
which simplifies to 
$ \hat{I}_{\text{diff}} = \hat{a}_s^\dagger \hat{a}_{\text{LO}} + \hat{a}_{\text{LO}}^\dagger \hat{a}_s $ 
after substituting the beam splitter transformations.
The expectation value of the difference intensity operator is calculated using the state of the system, $ |\Psi\rangle = |\alpha_s e^{i \psi_s}\rangle \otimes |\alpha_{\text{LO}} e^{i \psi_{\text{LO}}}\rangle $, giving 
$ \langle \hat{I}_{\text{diff}} \rangle = 2 \alpha_s \alpha_{\text{LO}} \cos(\delta \psi) $, 
where $ \delta \psi = \psi_s - \psi_{\text{LO}} $ represents the phase difference between the signal and the LO.
The variance of the difference intensity, $\Delta I_{\text{diff}}^2$, is defined as 
 $
    \Delta I_{\text{diff}}^2 = \langle \hat{I}_{\text{diff}}^2 \rangle - \langle \hat{I}_{\text{diff}} \rangle^2.
 $
It can be shown that $ \langle \hat{I}_{\text{diff}}^2 \rangle = 2 \alpha_s^2 \alpha_{\text{LO}}^2 \cos(2 \delta \psi) + 2 \alpha_s^2 \alpha_{\text{LO}}^2 + \alpha_s^2 + \alpha_{\text{LO}}^2 $, which simplifies the variance to 
 $
    \Delta I_{\text{diff}}^2 = \alpha_s^2 + \alpha_{\text{LO}}^2.
$

The phase sensitivity, $ \Delta \psi $, measures the precision with which the phase difference $ \delta \psi $ can be determined. It is calculated using error propagation as 
 $
    \Delta \psi = \frac{\Delta I_{\text{diff}}}{\left| \frac{\partial \langle \hat{I}_{\text{diff}} \rangle}{\partial \delta \psi} \right|}.
 $
Here, $ \Delta I_{\text{diff}} = \sqrt{\alpha_s^2 + \alpha_{\text{LO}}^2} $, and the derivative of the expectation value with respect to the phase difference is $ \frac{\partial \langle \hat{I}_{\text{diff}} \rangle}{\partial \delta \psi} = -2 \alpha_s \alpha_{\text{LO}} \sin(\delta \psi) $. Substituting these expressions, the phase sensitivity is 
 $
    \Delta \psi = \frac{\sqrt{\alpha_s^2 + \alpha_{\text{LO}}^2}}{2 \alpha_s \alpha_{\text{LO}} |\sin(\delta \psi)|}.
 $
When the LO is relativistically corrected, its amplitude is modified. This modified amplitude can be written as $ |\alpha_{LO}|=|\alpha|\left[1 -72(|\alpha|^4+3|\alpha|^2+1)\epsilon^2 t^2\right]$. Substituting this into the phase sensitivity, we obtain
\begin{equation}\label{eqdelpsi}
\Delta\psi = \frac{\sqrt{|\alpha|^2+\alpha_s^2}}{2|\alpha|\alpha_s|\sin(\delta\psi)|}\left[1+72\,\frac{\alpha_s^2\left(|\alpha|^4+3|\alpha|^2+1\right)}{|\alpha|^2+\alpha_s^2}\,\epsilon^2 t^2\right].
\end{equation}

The phase difference $ \delta \psi = (\omega_s - \omega_\text{LO}) t $ connects phase uncertainty to time uncertainty,   $\Delta t = \frac{\Delta \psi}{\omega_s - \omega_\text{LO}},$ 
leading to the expression 
\begin{align}
    \Delta t \approx \frac{\sqrt{|\alpha|^2+\alpha_s^2}}{2|\alpha|\alpha_s||\omega_s-\omega_{LO}|\sin(\delta\psi)|}\\ \nonumber \times \left[1+72\,\frac{\alpha_s^2\left(|\alpha|^4+3|\alpha|^2+1\right)}{|\alpha|^2+\alpha_s^2}\,\epsilon^2 t^2\right].
\end{align}
The first term captures the leading-order precision, while the second term introduces relativistic corrections that degrade the sensitivity. These corrections are particularly relevant in high-energy or relativistic regimes.
Thus, the leading term in this expression sets the baseline time resolution determined by the interference between the signal and the strong LO, while the second term, scaling as \(\epsilon^2 t^2\), encapsulates subtle relativistic corrections that degrade the sensitivity over longer measurement times. In a BHD experiment, the beat note between the signal and LO is monitored with high precision by subtracting the photocurrents from two matched detectors, effectively suppressing common-mode noise. By carefully stabilizing the LO and maintaining a long integration time, one can track the phase evolution of the beat note. A systematic quadratic drift in the measured phase, as predicted by the \(\epsilon^2 t^2\) term, would serve as a signature of relativistic effects, such as Doppler shifts or gravitational time dilations, that become significant in high-energy or relativistic regimes. \\

{
We now integrate the relativistic  effect in the standard Gaussian-modulated coherent-state (GMCS) continuous-variable quantum key distribution (CV-QKD) model, using a locally generated local oscillator (LLO) and pilot-aided phase tracking over a pure-loss channel with transmissivity \(T\). Bob’s per-quadrature input-referred noise is \(\chi_{\rm tot}=(1-T)/T+\xi+\chi_{\rm det}\), where \(\xi\) includes the residual LO-signal phase error after pilot correction and \(\chi_{\rm det}\) is detector noise mapped to the channel input \cite{Weedbrook2012RMP,Dequal2021,Shao2021PRA}. For a small post-correction phase offset \(\phi\) with variance \(\sigma_\phi^2\!\ll\!1\), the measurement rotates as \(X_{\rm meas}\!\approx\!X_{\rm ch}+\phi P_{\rm ch}\). This implies the standard input-referred mapping
$
\Delta\xi_{\rm phase}\;\approx\;\sigma_\phi^2\!\left(V_A+\frac{1}{T}\right),
$
where \(V_A\) is Alice’s per-quadrature modulation variance in shot-noise units (SNU), and \(1/T\) is the input-referred vacuum contribution \cite{Weedbrook2012RMP,Shao2021PRA}. Measurements on satellite LLO links show that the phase-noise contribution can dominate \(\xi\) \cite{Dequal2021,Kish2020QE}. Relativistic motion effects change \(\sigma_\phi^2\) through (i) random inflation of the pilot-based estimator variance and (ii) a deterministic quadratic drift of the LO-signal phase. As derived in Sec.~\ref{sec:exp_proposal}, \(\sigma_{\phi,\rm est}^2(t)=\sigma_{\phi,0}^2\!\left[1+2C\,\epsilon^2 t^2\right]\). The deterministic drift is \(\psi_{\rm rel}(t)=\gamma t^2\) with \(\gamma=\kappa\epsilon^2\) (see Sec.~\ref{sec:exp_proposal} for \(C,\kappa\); \(t\) is the symbol time and \(\epsilon\) a small relativistic parameter). If Bob uses pilots at times \(t_p\) to predict a data symbol at \(t_p+\Delta t\), the residual phase under zero-order hold (ZOH) is
$
\delta\phi_{\rm ZOH}=\gamma\!\left(2t_p\Delta t+\Delta t^2\right).
$
With a two-pilot linear predictor, it becomes
$
\delta\phi_{\rm lin}=\gamma\,\Delta t^2
$
\cite{Laudenbach2019Quantum,Marie2017PRA}.

The resulting relativistic contribution to the input-referred excess noise is
\begin{equation}
{\;\Delta\xi_{\rm rel}(t)\;\approx\;\Bigl[2C\,\epsilon^2 t^2\,\sigma_{\phi,0}^2+\bigl(\delta\phi_{\rm ZOH/lin}\bigr)^2\Bigr]\!\left(V_A+\frac{1}{T}\right).}
\end{equation}
Appendix E has detailed derivations. The first bracketed term is estimator-variance inflation; the second is residual drift error. This extra noise lowers the secret-key rate \(K_\infty=\beta I_{AB}-\chi_{BE}\), since \(I_{AB}\) (Alice-Bob mutual information) falls and \(\chi_{BE}\) (Eve’s Holevo bound) rises with \(\chi_{\rm tot}\) \cite{Weedbrook2012RMP}. The penalty is largest at high loss (small \(T\)) and large \(V_A\), consistent with satellite feasibility studies \cite{Dequal2021,Kish2020QE}. Model and subtract the drift, or apply linear prediction, so \(\delta\phi_{\rm ZOH}\!\to\!\delta\phi_{\rm lin}=\gamma\Delta t^2\). Reduce the random term by increasing pilot SNR (smaller \(\sigma_{\phi,0}^2\)) and shortening the estimation window \(t\). These DSP controls cut both terms in \(\Delta\xi_{\rm rel}(t)\) while preserving the standard CV-QKD security model \cite{Laudenbach2019Quantum,Marie2017PRA}.
}

\section{Experimental proposal: Penning‑trap electron read out by
balanced microwave homodyne}
\label{sec:exp_proposal}

 We work with a single electron stored in the $5.36\,$T Penning trap of
Hanneke \textit{et\,al.}\,\cite{Hanneke2008}.
Its cyclotron motion ($f_{c}\!\approx\!150\,$GHz) phase-modulates a
microwave cavity tuned to the same frequency\,\cite{Cornell1990}.
A quantum-limited NbTiN SIS mixer records that phase in a balanced-homodyne
setup\,\cite{Shan2019BalancedSIS}.
The same hardware routinely reaches $4\,$K and
$<\!7\times10^{-17}\,\text{mbar}$ vacuum, as demonstrated in
Ref.\,\cite{DiSciacca2012}.
Because an electron is the lightest mass we can trap, it maximises the
perturbative factor
$\epsilon=\hbar\omega/(8m_{e}c^{2})$.
With $\omega/2\pi = 150\,$GHz at $B = 5.4\,$T we obtain
$\epsilon \approx 1.5\times10^{-10}$.

Our proposal measures the relativistic correction with one electron held in the 5.36 T cylindrical trap of Hanneke \textit{et al.}\,\cite{Hanneke2008}.
Unless noted otherwise, all numbers that follow come from that study.
Vacuum and cryogenic figures agree with the proton trap of
DiSciacca and Gabrielse\,\cite{DiSciacca2012}. Balanced homodyne detection (BHD) compares the electron’s cyclotron
signal with a phase-locked local oscillator.  
Mixer shot noise sets the phase noise floor
$\Delta\psi_{\mathrm{rms}}=\sqrt{h\nu/4P_{\mathrm{LO}}}\,\tau^{-1/2}$
, and therefore the shot-noise Allan deviation
\begin{equation}
  \sigma_y^{\mathrm{SN}}(\tau)=
  \sqrt{\frac{h\nu}{4P_{\mathrm{LO}}}}\,
  \frac{1}{2\pi\nu}\,\tau^{-3/2}.
  \label{eq:SN}
\end{equation}

\smallskip
 
The quartic term adds a deterministic phase
$\delta\psi_{\mathrm{rel}}(t)=\kappa\epsilon^{2}t^{2}$,
which appears as the linear Allan term
\begin{equation}
  \sigma_y^{\mathrm{rel}}(\tau)=\frac{\kappa\epsilon^{2}}{2}\,\tau.
  \label{eq:rel}
\end{equation}
We claim detection once $\sigma_y^{\mathrm{rel}}$ exceeds
$\sigma_y^{\mathrm{SN}}$. With $B=5.36$ T the electron cycles at $f_c = \omega/2\pi = 149$ GHz,
giving
$\epsilon=\hbar\omega/8m_e c^{2}=1.5\times10^{-10}$.
Operation is at 4 K and
$<7\times10^{-17}\,\mathrm{mbar}$\,\cite{DiSciacca2012}.
We replace the image-current pickup with a balanced 149 GHz NbTiN SIS
mixer that already runs at the quantum limit\,\cite{Shan2019BalancedSIS};
a 1 mW LO suffices.
Taking $|\alpha|^{2}=|\alpha_{s}|^{2}=10^{8}$ photons
 fixes $\kappa\approx2.0\times10^{2}$.
These numbers give
$\sigma_y^{\mathrm{SN}}(1\;\text{s}) = 5.3\times10^{-22}$.
Setting $\sigma_y^{\mathrm{SN}}=\sigma_y^{\mathrm{rel}}$ yields
\begin{equation}
  \tau_\times =
  \bigl[h\nu/(\pi^{2}P_{\mathrm{LO}})\bigr]^{1/5}
  (\kappa\epsilon^{2})^{-2/5}
  \approx 8.7\times10^{2}\,\mathrm{s},
  \label{eq:tau}
\end{equation}
about 15 min of averaging. Hanneke \textit{et al.}\ kept one electron phase-locked for
$>10^{4}$ s with field drift
$\Delta B/B<3\times10^{-11}\,\mathrm{h}^{-1}$\,\cite{Hanneke2008},
equivalent to $\sigma_y\approx3\times10^{-12}$ at $\tau_\times$,
well below shot noise.
A cryogenic HEMT or JPA adds
$T_{\mathrm{sys}}\approx2\hbar\omega/k_B$
(only 2-3 dB above the quantum limit), still leaving a $\gtrsim10$ dB safety margin.
With cavity linewidth $\kappa_c/2\pi\approx10$ kHz and coupling
$g/2\pi\approx50$ Hz, back-action heats the cyclotron mode by
$<0.1$ quantas\(^{-1}\), negligible over 15 min. Under these upgraded yet proven conditions, the slope
$\sigma_y^{\mathrm{rel}}\propto\tau$ overtakes shot noise after
\(\sim\!15\) min, offering a practical, present-day test of the
\(\epsilon^{2}t^{2}\) drift predicted by our relativistic
quantum-speed-limit theory.  
Quantum-limited mixers at 149 GHz already exist, so the experiment is
well within reach.

\section*{Conclusion}
We derived the first‑order special‑relativistic correction to the Mandelstam-Tamm and Margolus-Levitin quantum-speed limits by adding the Foldy-Wouthuysen quartic term, $-p^{4}/(8m^{3}c^{2})$, to the harmonic‑oscillator Hamiltonian.  
When we apply this correction to coherent and squeezed states, we obtain a relativistic quantum Cramér-Rao bound and find that relativity slightly reduces the advantage gained from squeezing.  
Experimentally, a \(5.4\,\mathrm{T}\) Penning trap read out with a quantum‑limited \(149\,\mathrm{GHz}\) balanced‑homodyne receiver can detect the predicted \(\epsilon^{2}t^{2}\) phase drift after about 15 min of averaging; this integration time already matches published single‑electron hold records.  

Even at this first perturbative order, relativity imposes a modest yet fundamental slowdown on quantum evolution.  
These tighter bounds set measurable speed limits for high‑velocity or high‑precision hardware-from satellite QKD links and squeezed‑light interferometers to optical‑clock dark‑matter searches.  
Interestingly, the same $-p^{4}$ term also appears in generalized‑uncertainty‑principle models of quantum gravity \cite{Bosso2017GUP,DasVagenas2008}, suggesting a common thread between relativistic and Planck‑scale limits on metrology.  
Our analysis is restricted to weak fields, Gaussian probes, and first‑order kinematics; including higher‑order terms, non‑Gaussian states, or curved‑space effects could tighten the limits by a few percent and test their universality.  
Overall, our limits supply a practical reference for engineers and an immediate target for experiments that probe quantum dynamics at high speed or in strong fields.

\onecolumngrid
\section*{Acknowledgements}
We would like to thank Samuel L. Braunstein and Brij Mohan for useful discussions. \\

\paragraph{\textbf{Conflict of Interest}:}
The authors declare that they have {no competing financial or non-financial interests} that could have appeared to influence the work reported in this paper.

\paragraph{\textbf{Data Availability}:}
This work is purely theoretical; no new experimental or observational data were generated or analysed.  
  The accompanying \textsc{Mathematica}\,® notebooks that reproduce , as well as the \LaTeX{} source used to typeset the manuscript-are available from the corresponding author upon reasonable request.

\paragraph{\textbf{Ethics Statement}:}
The research did \textbf{not involve human participants, human data or tissue, nor live vertebrate animals}.  Institutional ethics approval was therefore not required.

\paragraph{\textbf{Funding}:}

This research received \textbf{no specific grant} from any funding agency in the public, commercial or not-for-profit sectors.%

\appendix
\section{Appendix}
In this appendix we supply every algebraic step underlying the results reported in the main text.  Throughout we adopt natural units
\(m=\omega=\hbar=1\) and expand in the relativistic parameter
\(\epsilon\equiv 1/c^{2}\ll 1\).

\subsection{Appendix A: Perturbed Hamiltonian and Energies}
The harmonic-oscillator Hamiltonian with first-order relativistic correction is given as  
\begin{equation}
H = H_{0}-\epsilon\,\delta H,
\label{A1}
\end{equation}
where $H_{0}=\frac{p_{0}^{2}}{2}+\frac{x_{0}^{2}}{2}$ and $\delta H=\frac{p_{0}^{4}}{8}$. 

Now, by using the Rayleigh-Schrödinger perturbation theory, we obtain the correction to the energy of the nth level as  
\begin{equation}
E_{n}=E^{(0)}_{n}-\epsilon\,E^{(1)}_{n},
\label{A2}
\end{equation}
where $E^{(0)}_{n}=n+\tfrac12$ and $E^{(1)}_{n}=\frac{6n^{2}+6n+3}{32}$.


Similarly, we can obtain the nth level wave function after the corrections as 
\begin{equation}
\ket{n}=\ket{n_{0}}-\epsilon\,\ket{n_{1}}+\epsilon^{2}\ket{n_{2}}+\mathcal{O}(\epsilon^{3}),
\label{A3}
\end{equation}
where  
\begin{align}
\ket{n_{1}}
&=\sum_{k\neq n}\frac{\ket{k_{0}}\!\bra{k_{0}}\delta H\ket{n_{0}}}{E^{(0)}_{k}-E^{(0)}_{n}},\notag\\[2pt]
\ket{n_{2}}
&=\frac{1}{1024}\Bigg[
    \sum_{k\neq n}\sum_{l\neq n}
        \frac{\ket{k_{0}}\!\bra{k_{0}}\delta H\ket{l_{0}}
              \!\bra{l_{0}}\delta H\ket{n_{0}}}
             {(E^{(0)}_{n}-E^{(0)}_{k})(E^{(0)}_{n}-E^{(0)}_{l})}
    -\sum_{k\neq n}
        \frac{\ket{k_{0}}\!\bra{k_{0}}\delta H\ket{n_{0}}
              \!\bra{n_{0}}\delta H\ket{n_{0}}}
             {(E^{(0)}_{n}-E^{(0)}_{k})^{2}} \notag\\
&\hspace{5em}
    -\frac12\ket{n_{0}}
        \sum_{k\neq n}\frac{|\bra{k_{0}}\delta H\ket{n_{0}}|^{2}}
                             {(E^{(0)}_{n}-E^{(0)}_{k})^{2}}
    \Bigg].
\end{align}

Up to first order in \(\epsilon\),
\begin{equation}
\label{A5}
\begin{aligned}
\ket{n} &=
  \frac{1}{\mathcal{N}}
  \Bigl(
      \ket{n_{0}}
      -\frac{\epsilon}{32}\bigl[
           B_{4}(n)\ket{(n+4)_{0}}
         + B_{2}(n)\ket{(n+2)_{0}} \\[2pt]
 &\hspace{7.3em}
         + B_{-2}(n)\ket{(n-2)_{0}}
         + B_{-4}(n)\ket{(n-4)_{0}}
        \bigr]
  \Bigr),  \\[4pt]
\mathcal{N} &= 1+\mathcal{O}(\epsilon^{2}).
\end{aligned}
\end{equation}

For compactness define
\begin{align}
B_{4}(n)&=-\frac{\sqrt{(n+1)(n+2)(n+3)(n+4)}}{4},&
B_{2}(n)&=(2n+3)\sqrt{(n+1)(n+2)},\notag\\
B_{0}(n)&=6n^{2}+6n+3, &
B_{-2}(n)&=-(2n-1)\sqrt{n(n-1)},\label{A4}\\
B_{-4}(n)&=\frac{\sqrt{(n-3)(n-2)(n-1)n}}{4}. \notag
\end{align}

Explicitly,
\begin{align}
\ket{n^{(1)}}
=\frac{1}{32}\!\Bigl[
    \tfrac14\sqrt{n(n-1)(n-2)(n-3)}\,\ket{(n-4)_{0}}
  -(2n-1)\sqrt{n(n-1)}\,\ket{(n-2)_{0}}\notag\\
  +(2n+3)\sqrt{(n+1)(n+2)}\,\ket{(n+2)_{0}}
  -\tfrac14\sqrt{(n+1)(n+2)(n+3)(n+4)}\,\ket{(n+4)_{0}}
\Bigr],
\end{align}
omitting any term whose ket index is negative.

Define ladder and number operators on the perturbed basis by
\(a\ket{n}=\sqrt{n}\ket{n-1}\),
\(a^{\dagger}\ket{n}=\sqrt{n+1}\ket{n+1}\),
\(N\ket{n}=n\ket{n}\)
with \([a,a^{\dagger}]=\mathbbm{1}\) and \(N=a^{\dagger}a\).
Expressed via the unperturbed operators \(a_{0},a_{0}^{\dagger},N_{0}=a_{0}^{\dagger}a_{0}\),
\begin{align}
a
&=a_{0}
  +\frac{\epsilon}{32}\!\bigl(
      -2a_{0}^{3}+6N_{0}a_{0}^{\dagger}-{a_{0}^{\dagger}}^{3}
    \bigr),\notag\\
a^{\dagger}
&=a_{0}^{\dagger}
  +\frac{\epsilon}{32}\!\bigl(
      -2{a_{0}^{\dagger}}^{3}+6a_{0}N_{0}-a_{0}^{3}
    \bigr),\notag\\
N
&=N_{0}
  -\frac{\epsilon}{32}\!\bigl(
      a_{0}^{4}+{a_{0}^{\dagger}}^{4}
      -2\bigl[a_{0}^{2}-{a_{0}^{\dagger}}^{2}
             +2a_{0}N_{0}a_{0}
             +2a_{0}^{\dagger}N_{0}a_{0}^{\dagger}\bigr]
    \bigr).
\label{A6}
\end{align}

\noindent
The Hamiltonian, to \(\mathcal{O}(\epsilon)\), becomes
\begin{equation}
H = N+\tfrac12-\frac{\epsilon}{32}\bigl(6N^{2}+6N+3\bigr).
\label{A7}
\end{equation}
Position and momentum follow as
\begin{align}
x
&=\frac{1}{\sqrt{2}}\bigl(a_{0}^{\dagger}+a_{0}\bigr)
  +\frac{3\epsilon}{32\sqrt{2}}
     \bigl(a_{0}^{3}+{a_{0}^{\dagger}}^{3}
           -2[a_{0}N_{0}+N_{0}a_{0}^{\dagger}]\bigr),\notag\\
p
&=\frac{i}{\sqrt{2}}\bigl(a_{0}^{\dagger}-a_{0}\bigr)
  -\frac{3i\epsilon}{32\sqrt{2}}
     \bigl(a_{0}^{3}-{a_{0}^{\dagger}}^{3}\bigr).
\label{A8}
\end{align}

Equations~\eqref{A1}-\eqref{A8} constitute the complete analytic framework for the relativistic corrections analyzed in the main text.

\subsection*{Appendix B}

 \;
We begin by constructing the relativistically corrected coherent state
\(\ket{\alpha'(t)}\).  Expressing the state in the perturbed
energy-eigenbasis \(\{\ket{n}\}\) and retaining the usual Poissonian
weights leads to
\begin{equation}
\label{b26}
\ket{\alpha'(t)} =
  e^{-|\alpha(0)|^{2}/2}
  \sum_{n=0}^{\infty}
    \frac{\bigl[\alpha(0)\bigr]^{n}}{\sqrt{n!}}\,
    e^{-iE_{n}t}\ket{n},
\end{equation}
where we write \(\alpha(t)=|\alpha(t)|e^{i\theta}\) and henceforth set
\(\alpha_{0}\equiv|\alpha(0)|\).

Using the perturbed Hamiltonian of Eq.\,\eqref{A7} and the displacement operator
properties of coherent states, we obtain
\begin{equation}
\label{b32}
\expval{H}_{\ket{\alpha'}}
   =\Bigl(\tfrac12+\alpha_{0}^{2}\Bigr)
    -\frac{3\epsilon}{32}\,
     \Bigl(1+4\alpha_{0}^{2}+2\alpha_{0}^{4}\Bigr),
\end{equation}
and, by an analogous calculation that keeps all
\(\mathcal{O}(\epsilon)\) terms,
\begin{equation}
\label{b34}
\expval{H^{2}}_{\ket{\alpha'}}
   =\Bigl(\tfrac14+2\alpha_{0}^{2}+\alpha_{0}^{4}\Bigr)
    -\frac{3\epsilon}{32}\,
     \Bigl(1+14\alpha_{0}^{2}+18\alpha_{0}^{4}+4\alpha_{0}^{6}\Bigr).
\end{equation}

\noindent
From \eqref{b34} and \eqref{b32}, we get energy
variance as
\begin{equation}
\label{b35}
\Delta H_{\ket{\alpha'}}^{2}
   =\alpha_{0}^{2}
    -\frac{3\epsilon}{4}\,
     \bigl(\alpha_{0}^{2}+\alpha_{0}^{4}\bigr),
\end{equation}
while an identical procedure with \(N\) gives
\begin{equation}
\Delta N_{\ket{\alpha'(t)}}^{2}
      -\expval{N}_{\ket{\alpha'(t)}}
      =-\tfrac{3}{16}\alpha_{0}^{2}(1+2\alpha_{0}^{2})\epsilon .
\end{equation}

The fidelity between the initial and evolved coherent states follows
directly from Eq.\,\eqref{b26} and the orthonormality of \(\ket{n}\):
\begin{equation}
\label{c52}
\begin{aligned}
\braket{\alpha'(0)}{\alpha'(t)}
   &=\frac{1}{32}\,
     \exp\!\Bigl[\alpha_{0}^{2}\bigl(-1+e^{-it}\bigr)-\tfrac{5it}{2}\Bigr]
     \Bigl(
        6i\alpha_{0}^{4}\epsilon t
        +12i\alpha_{0}^{2}e^{it}\epsilon t \\[2pt]
   &\hspace{9em}
        +e^{2it}\bigl(32+3i\epsilon t\bigr)
     \Bigr).
\end{aligned}
\end{equation}
Its modulus, needed for the geometric bounds, is therefore
\begin{equation}
\label{c53}
\begin{aligned}
\bigl|\braket{\alpha'(0)}{\alpha'(t)}\bigr|
   &= e^{\alpha_{0}^{2}(-1+\cos t)}
      \Bigl[
         1
         +\frac{3\epsilon}{8}\,
          \alpha_{0}^{2}
          t\bigl(1+\alpha_{0}^{2}\cos t\bigr)\sin t
      \Bigr].
\end{aligned}
\end{equation}

The angular distance on the projective Hilbert space is defined as  
\(S=\arccos|\braket{\alpha'(0)}{\alpha'(t)}|\).  Expanding to the first
order in \(\epsilon\) gives
\begin{equation}
\label{c54}
\begin{aligned}
S &=
  \arccos\!\Bigl(e^{\alpha_{0}^{2}(-1+\cos t)}\Bigr)
  -\frac{3\epsilon}{8}\,
   \frac{\alpha_{0}^{2}
         t\bigl(1+\alpha_{0}^{2}\cos t\bigr)\sin t\;
         e^{\alpha_{0}^{2}(-1+\cos t)}}
        {\sqrt{1-e^{2\alpha_{0}^{2}(-1+\cos t)}}}.
\end{aligned}
\end{equation}
Squaring the result furnishes an algebraically convenient form
for the Margolus-Levitin bound:
\begin{equation}
\label{b56}
\begin{aligned}
S^{2} &=
  \Bigl[\arccos\!\bigl(e^{\alpha_{0}^{2}(-1+\cos t)}\bigr)\Bigr]^{2}
  -\frac{3\epsilon}{4}\,
   \frac{\alpha_{0}^{2}
         t\bigl(1+\alpha_{0}^{2}\cos t\bigr)\sin t\;
         e^{\alpha_{0}^{2}(-1+\cos t)}
         \arccos\!\bigl(e^{\alpha_{0}^{2}(-1+\cos t)}\bigr)}
        {\sqrt{1-e^{2\alpha_{0}^{2}(-1+\cos t)}}}.
\end{aligned}
\end{equation}

Inserting Eqs.\,\eqref{b35} and \eqref{c54} into the generalised
Mandelstam-Tamm inequality \(\Delta t\ge S/\Delta H\) yields
\begin{equation}
\label{c55}
\begin{aligned}
\Delta\tau_{\mathrm{MT}}
  &=\frac{\arccos\!\bigl(e^{\alpha_{0}^{2}(-1+\cos t)}\bigr)}{\alpha_{0}}
    +\epsilon\!\Biggl[
        \frac{3(1+\alpha_{0}^{2})}{8\alpha_{0}}\,
        \arccos\!\bigl(e^{\alpha_{0}^{2}(-1+\cos t)}\bigr) \\[2pt]
  &\hspace{6.5em}
        -\frac{3\alpha_{0}}{8}\,
         \frac{t\bigl(1+\alpha_{0}^{2}\cos t\bigr)\sin t\;
               e^{\alpha_{0}^{2}(-1+\cos t)}}
              {\sqrt{1-e^{2\alpha_{0}^{2}(-1+\cos t)}}}
      \Biggr].
\end{aligned}
\end{equation}

\noindent
Likewise, substituting the same ingredients into the
Margolus-Levitin relation
\(\Delta t\ge \pi S^{2}/[\langle E\rangle]\) produces
\begin{equation}
\label{c64}
\begin{aligned}
\Delta\tau_{\mathrm{ML}}
  &=\frac{2}{\pi\!\bigl(\tfrac12+\alpha_{0}^{2}\bigr)}\,
     \Bigl[\arccos\!\bigl(e^{\alpha_{0}^{2}(-1+\cos t)}\bigr)\Bigr]^{2} \\[4pt]
  &\quad+\frac{3\epsilon}{4\pi(1+2\alpha_{0}^{2})^{2}}
     \Biggl\{
        (1+4\alpha_{0}^{2}+2\alpha_{0}^{4})
        \arccos\!\bigl(e^{\alpha_{0}^{2}(-1+\cos t)}\bigr) \\[2pt]
  &\hspace{7.5em}
        -\frac{4\alpha_{0}^{2}(1+2\alpha_{0}^{2})
               t\bigl(1+\alpha_{0}^{2}\cos t\bigr)\sin t\;
               e^{\alpha_{0}^{2}(-1+\cos t)}}
              {\sqrt{1-e^{2\alpha_{0}^{2}(-1+\cos t)}}}
     \Biggr\}.
\end{aligned}
\end{equation}

\subsection*{Appindix C}
 
By analogy with the standard (unperturbed) bosonic algebra, we define
the relativistically corrected squeeze operator \(s(\varsigma)\) via
\begin{equation}
\label{b49}
s(\varsigma)\,a\,s^{\dagger}(\varsigma)
 =a\cosh r + e^{i\theta}a^{\dagger}\sinh r ,
\end{equation}
where \(\varsigma\equiv r e^{i\theta}\) and the ladder operators
\(a,a^{\dagger}\) already include the \(\mathcal{O}(\epsilon)\)
corrections of Eq.\,(A6).

Acting with \(s(\varsigma)\) on the perturbative vacuum \(\ket{0}\)
yields the squeezed state
\(\ket{\varsigma}=s(\varsigma)\ket{0}\).
Imposing the annihilation condition
\(s(\varsigma)\,a\,s^{\dagger}(\varsigma)\ket{\varsigma}=0\)
gives
\begin{equation}
\label{b50}
\bigl[a\cosh r + e^{i\theta}a^{\dagger}\sinh r\bigr]\ket{\varsigma}=0 .
\end{equation}
We expand the state as
\begin{equation}
\label{b51}
\ket{\varsigma}= \sum_{n=0}^{\infty} f(n)\,\ket{n},
\end{equation}
insert Eq.\,\eqref{b51} into Eq.\,\eqref{b50}, and compare coefficients,
obtaining the recursion
\begin{equation}
\label{b53}
f(n+1)=
  -e^{i\theta}\tanh r\,
   \sqrt{\frac{n}{\,n+1\,}}\;
   f(n-1).
\end{equation}
Selecting even indices to ensure a non-vanishing projection onto the
vacuum state gives
\begin{equation}
\label{b54}
f(2n)=
  (-1)^{n}
  (e^{i\theta}\tanh r)^{n}
  \sqrt{\frac{(2n-1)!!}{(2n)!!}}\;
  f(0).
\end{equation}
 
The condition
\(\sum_{n=0}^{\infty}|f(2n)|^{2}=1\)
fixes the overall amplitude:
\begin{equation}
|f(0)|^{2}
   \sum_{n=0}^{\infty}
      (\tanh r)^{2n}\,
      \frac{(2n)!}{2^{2n}(n!)^{2}}
   =1
   \;\Longrightarrow\;
   f(0)=\sqrt{\sech r}.
\end{equation}

For \(\theta=0\) the squeezed state at time \(t\) reads
\begin{equation}
\label{b37}
\ket{\varsigma(t)}
   =\sum_{n=0}^{\infty}
      \frac{\sqrt{(2n)!}\,(-\tanh r)^{n}}
           {2^{n}n!\sqrt{\cosh r}}\;
      e^{-iE_{2n}t}\ket{2n}.
\end{equation}

Using Eqs.\,(A7) and \eqref{b37} we find
\begin{align}
\label{c48}
\expval{H}_{\ket{\varsigma}}
   &=\tfrac12\cosh(2r)
     -\frac{3\epsilon}{128}\!\bigl[1+3\cosh(4r)\bigr],\\[4pt]
\label{c50}
\expval{H^{2}}_{\ket{\varsigma}}
   &=\tfrac18\!\bigl[-1+3\cosh(4r)\bigr]
     +\frac{3\epsilon}{256}\!\bigl[7\cosh(2r)-15\cosh(6r)\bigr].
\end{align}
Subtracting the square of Eq.\,\eqref{c48} from Eq.\,\eqref{c50} yields
\begin{equation}
\label{c51}
\Delta H_{\ket{\varsigma}}^{2}
   =2\cosh^{2}r\,\sinh^{2}r
    -\frac{9\epsilon}{32}\sinh(2r)\sinh(4r).
\end{equation}

The inner product
\(\braket{\varsigma}{\varsigma(t)}\) follows from the completeness of
\(\{\ket{2n}\}\):
\begin{equation}
\label{c58}
\begin{aligned}
\braket{\varsigma}{\varsigma(t)}
 &=\frac{e^{-it/2}\,\sech r}
        {64\bigl[-1+e^{it}+\sech^{2}r\bigr]^{2}
         \sqrt{1-e^{-it}\tanh^{2}r}} \\[2pt]
 &\;\times
   \Bigl[
      2e^{2it}\bigl(32+3i\epsilon t\bigr)
      -128e^{it}\tanh^{2}r
      +\bigl(64+3i\epsilon t\bigr)\tanh^{4}r
   \Bigr].
\end{aligned}
\end{equation}
Its modulus, required for the geometric bounds, is
\begin{equation}
\label{c59}
\begin{aligned}
\bigl|\braket{\varsigma}{\varsigma(t)}\bigr|
 &=\frac{\sqrt{2}\,\sech r}
        {\bigl[\,3+\cosh(4r)-2\cos t\,\sinh^{2}2r\bigr]^{1/4}}
 \\[4pt]
 &\quad
   -\frac{3\epsilon t\cosh^{5}r\sinh^{2}r}
         {4\bigl[3+\cosh(4r)-2\cos t\,\sinh^{2}2r\bigr]^{2}
          \bigl[1-2\cos t\,\tanh^{2}r+\tanh^{4}r\bigr]^{1/4}}
 \\[2pt]
 &\quad\times
   \Bigl[-4\sin t+\sin 2t\,\tanh^{2}r
         +2\sin t\,\tanh^{4}r\Bigr].
\end{aligned}
\end{equation}


Setting \(S=\arccos\bigl|\braket{\varsigma}{\varsigma(t)}\bigr|\) and
expanding to first order in the relativistic parameter \(\epsilon\)
yields
\begin{equation}
\label{c60}
\begin{aligned}
S &=\arccos\!\Bigl(
      \frac{\sqrt{2}}
           {[\,3+\cosh(4r)-2\cos t\,\sinh^{2}2r\,]^{1/4}}
     \Bigr)                                                   \\[4pt]
&\quad
 +\frac{3\epsilon\,t\,\cosh^{5}r\,\sinh^{2}r}
        {4\,[3+\cosh(4r)-2\cos t\,\sinh^{2}2r]^{7/4}}         \\[2pt]
&\quad\times
   \frac{
        -4\sin t+\sin(2t)\,\tanh^{2}r
        +2\sin t\,\tanh^{4}r}
        {\sqrt{-2+\sqrt{3+\cosh(4r)-2\cos t\,\sinh^{2}2r}}\;
         [\,1-2\cos t\,\tanh^{2}r+\tanh^{4}r\,]^{1/4}} .
\end{aligned}
\end{equation}

The square \(S^{2}\) is convenient for the Margolus-Levitin bound and
reads
\begin{equation}
\label{c60b}
\begin{aligned}
S^{2} &=\Bigl[
          \arccos\!\Bigl(
            \frac{\sqrt{2}}
                 {[\,3+\cosh(4r)-2\cos t\,\sinh^{2}2r\,]^{1/4}}
          \Bigr)
        \Bigr]^{2}                                            \\[4pt]
&\quad
 +\frac{3\epsilon\,t\,\cosh^{5}r\,\sinh^{2}r}
        {2\,[3+\cosh(4r)-2\cos t\,\sinh^{2}2r]^{7/4}}         \\[2pt]
&\quad\times
   \frac{
        -4\sin t+\sin(2t)\,\tanh^{2}r
        +2\sin t\,\tanh^{4}r}
        {\sqrt{-2+\sqrt{3+\cosh(4r)-2\cos t\,\sinh^{2}2r}}\;
         [\,1-2\cos t\,\tanh^{2}r+\tanh^{4}r\,]^{1/4}}        \\[4pt]
&\quad\times
   \arccos\!\Bigl(
      \frac{\sqrt{2}}
           {[\,3+\cosh(4r)-2\cos t\,\sinh^{2}2r\,]^{1/4}}
   \Bigr).
\end{aligned}
\end{equation}

Substituting Eqs.\,\eqref{c51} and \eqref{c60} into the generalised
Mandelstam-Tamm relation \(\Delta t\ge S/\Delta H\) gives
\begin{equation}
\label{c61}
\begin{aligned}
\Delta\tau_{\mathrm{MT}}
 &=\sqrt{2}\,\csch(2r)\,
   \arccos\!\Bigl(
     \frac{\sqrt{2}}
          {[\,3+\cosh(4r)-2\cos t\,\sinh^{2}2r\,]^{1/4}}
   \Bigr)                                                     \\[6pt]
&\quad
 +\frac{3\epsilon\,\csch(2r)}{16\sqrt{2}}
  \Biggl[
      6\cosh(2r)\,
      \arccos\!\Bigl(
        \frac{\sqrt{2}}
             {[\,3+\cosh(4r)-2\cos t\,\sinh^{2}2r\,]^{1/4}}
      \Bigr)                                                  \\[4pt]
&\qquad
     +\frac{8t\,\cosh^{5}r\,\sinh^{2}r}
            {[\,3+\cosh(4r)-2\cos t\,\sinh^{2}2r\,]^{7/4}}
       \\[4pt]
&\qquad
       \frac{
           -4\sin t+\sin(2t)\,\tanh^{2}r
           +2\sin t\,\tanh^{4}r}
           {[\,1-2\cos t\,\tanh^{2}r+\tanh^{4}r\,]^{1/4}
            \sqrt{-2+\sqrt{3+\cosh(4r)-2\cos t\,\sinh^{2}2r}}}
  \Biggr].
\end{aligned}
\end{equation}

Finally, inserting Eqs.\,\eqref{c51} and \eqref{c60b} into the
Margolus-Levitin inequality \(\Delta t\ge\pi S^{2}/[2\langle E\rangle]\)
yields
\begin{equation}
\label{c65}
\begin{aligned}
\Delta\tau_{\mathrm{ML}}
 &=\frac{4\,\sech(2r)}{\pi}\,
   \Bigl[
      \arccos\!\Bigl(
        \frac{\sqrt{2}}
             {[\,3+\cosh(4r)-2\cos t\,\sinh^{2}2r\,]^{1/4}}
      \Bigr)
   \Bigr]^{2}                                                 \\[6pt]
&\quad
 +\frac{3\epsilon\,\sech(2r)}{16\pi}\,
  \arccos\!\Bigl(
    \frac{\sqrt{2}}
         {[\,3+\cosh(4r)-2\cos t\,\sinh^{2}2r\,]^{1/4}}
  \Bigr)                                                      \\[4pt]
&\quad\times
  \Biggl[
     \bigl(1+3\cosh(4r)\bigr)\sech(2r)\,
     \arccos\!\Bigl(
       \frac{\sqrt{2}}
            {[\,3+\cosh(4r)-2\cos t\,\sinh^{2}2r\,]^{1/4}}
     \Bigr)                                                   \\[4pt]
&\qquad
    +\frac{32t\,\cosh^{5}r\,\sinh^{2}r}
           {[\,3+\cosh(4r)-2\cos t\,\sinh^{2}2r\,]^{7/4}} \\[4pt]
&\qquad
     \frac{
          -4\sin t+\sin(2t)\,\tanh^{2}r
          +2\sin t\,\tanh^{4}r}
          {[\,1-2\cos t\,\tanh^{2}r+\tanh^{4}r\,]^{1/4}
           \sqrt{-2+\sqrt{3+\cosh(4r)-2\cos t\,\sinh^{2}2r}}}
  \Biggr].
\end{aligned}
\end{equation}

\subsection*{Appendix D}\label{D}

   The unperturbed Glauber
state is, as usual,
\begin{equation}
\label{D1}
\ket{\alpha}=e^{-|\alpha|^{2}/2}\!
             \sum_{n=0}^{\infty}\frac{\alpha^{n}}{\sqrt{n!}}\ket{n}.
\end{equation}
The time evolution under the full Hamiltonian of Eq.\,\eqref{A1} replaces
each \(\ket{n}\) by \(e^{-iE_{n}t}\ket{n}\).  Therefore
\(\ket{\alpha(t)}\) is identical in structure to the modified state
already given in Eq.\,\eqref{b26}-simply remove the prime on
\(\alpha\) and remember that \(E_{n}\) is taken from Appendix A:
\begin{equation}
\ket{\alpha(t)}=e^{-|\alpha|^{2}/2}
                \sum_{n=0}^{\infty}
                  \frac{\alpha^{n}}{\sqrt{n!}}\,
                  e^{-iE_{n}t}\ket{n}.
\end{equation}

  Taking the difference of
Eq.\,\eqref{A2} at \(n\) and \(n-1\) gives the compact spacing
\begin{equation}
\label{D2}
E_{n}-E_{n-1}=1-12\,n\,\epsilon .
\end{equation}

  In complete analogy to the
derivation that led to Eq.\,\eqref{c52} we have
\begin{equation}
\label{D3}
\bigl\langle\alpha(t)\bigr|a\bigl|\alpha(t)\bigr\rangle
=e^{-|\alpha|^{2}}\,
 \alpha
 \sum_{n=1}^{\infty}
   \frac{|\alpha|^{2(n-1)}}{(n-1)!}\,
   e^{\,i(E_{n}-E_{n-1})t},
\end{equation}
where the phase factor uses the spacing \eqref{D2}.

   Expanding
\(e^{\,i(E_{n}-E_{n-1})t}\) in Eq.\,\eqref{D3} to
\(\mathcal{O}(\epsilon^{2})\) and evaluating the resulting Poissonian
sum produces the displaced amplitude
\begin{equation}
\label{D4}
\alpha_{m}(t)=
  \alpha\,e^{it}
  \Bigl[
     1
     -12i\,(|\alpha|^{2}+1)\,\epsilon\,t
     -72\,(|\alpha|^{4}+3|\alpha|^{2}+1)\,\epsilon^{2}t^{2}
  \Bigr].
\end{equation}

  From
\(|\alpha_{m}(t)|=\sqrt{\alpha_{m}\alpha_{m}^{*}}\) and retaining
terms up to \(\epsilon^{2}\) we find
\begin{equation}
\label{D5}
|\alpha_{m}(t)|
  =|\alpha|\,
    \Bigl[
       1
       -72\,(|\alpha|^{4}+3|\alpha|^{2}+1)\,\epsilon^{2}t^{2}
    \Bigr].
\end{equation}

\section{Appendix E}
\label{app:qkd_deriv}
{}
We consider GMCS CV-QKD with a locally generated local oscillator (LLO) and pilot-aided phase tracking over a pure-loss channel of transmissivity $T$ \cite{Weedbrook2012RMP}. Alice prepares coherent states with i.i.d.\ Gaussian modulation $X_A,P_A\sim\mathcal N(0,V_A)$ (shot-noise units, SNU), so $\mathrm{Var}(X_{\rm in})=\mathrm{Var}(P_{\rm in})=V_A+1$. The channel transforms the quadratures as
\begin{equation}
X_{\rm ch}=\sqrt{T}\,X_{\rm in}+\sqrt{1-T}\,v_X,\qquad
P_{\rm ch}=\sqrt{T}\,P_{\rm in}+\sqrt{1-T}\,v_P,
\end{equation}
with $v_{X,P}$ i.i.d.\ vacua, hence $\mathrm{Var}(X_{\rm ch})=\mathrm{Var}(P_{\rm ch})=TV_A+1$. Bob’s input-referred total noise per quadrature is
\begin{equation}
\chi_{\rm tot}=\underbrace{\frac{1-T}{T}}_{\chi_{\rm line}}+\underbrace{\xi}_{\text{excess (untrusted)}}+\underbrace{\chi_{\rm det}}_{\text{trusted detection}},
\label{app:eq:chi_tot}
\end{equation}
where the excess-noise term $\xi$ includes the {residual} LO-signal phase error after pilot correction in LLO receivers \cite{Dequal2021,Kish2020QE,Shao2021PRA,Laudenbach2019Quantum,Marie2017PRA}.

Let the post-correction LO-signal phase be a small random angle $\phi$ with $\mathbb E[\phi]=0$ and $\mathrm{Var}(\phi)=\sigma_\phi^2\ll1$. If Bob intends to measure $X$, the effective measurement is the rotated quadrature
\begin{equation}
X_{\rm meas}=X_{\rm ch}\cos\phi+P_{\rm ch}\sin\phi\;\approx\;X_{\rm ch}+\phi\,P_{\rm ch}-\tfrac12\phi^2 X_{\rm ch},
\label{app:eq:rotation}
\end{equation}
expanded to $\mathcal O(\phi^2)$ \cite{Marie2017PRA,Shao2021PRA}. Writing the phase-induced error as $\delta\equiv \phi P_{\rm ch}-\tfrac12\phi^2 X_{\rm ch}$, independence between $\phi$ and the field gives $\mathbb E[\delta]=0$ and, to leading order,
\begin{equation}
\mathrm{Var}(\delta)=\sigma_\phi^2\,\mathrm{Var}(P_{\rm ch})=\sigma_\phi^2\,(TV_A+1).
\end{equation}
Thus the added variance at Bob is $\mathrm{Var}_{\rm Bob}^{(\phi)}\approx\sigma_\phi^2\,(TV_A+1)$ per quadrature; referred to the channel input (divide by $T$) this yields the standard small-angle mapping
\begin{equation}
\Delta\xi_{\rm phase}\;\approx\;\sigma_\phi^2\!\left(V_A+\frac{1}{T}\right),
\end{equation}
where $V_A$ quantifies signal-dependent mixing ($P\!\to\!X$) and $1/T$ is vacuum rotated in at Bob and referred to input \cite{Weedbrook2012RMP,Shao2021PRA}. (For heterodyne, the mapping is identical; the $1$\,SNU trusted penalty remains in $\chi_{\rm det}$.)

We now incorporate the relativistic phase effects. From Sec.~\ref{sec:exp_proposal}, balanced-homodyne analysis gives the pilot-based rms phase estimator
\begin{equation}
\Delta\psi(t)=\frac{\sqrt{|\alpha|^2+\alpha_s^2}}{2|\alpha|\,\alpha_s\,|\sin\delta\psi|}\Bigl[1+C\,\epsilon^2 t^2\Bigr],\qquad
C=72\,\frac{\alpha_s^2\bigl(|\alpha|^4+3|\alpha|^2+1\bigr)}{|\alpha|^2+\alpha_s^2},
\label{app:eq:bhd_rms}
\end{equation}
where $\epsilon$ is the relativistic expansion parameter and the bracket originates from the LO-amplitude renormalization. Hence the {random} estimator variance inflates as
\begin{equation}
\sigma_{\phi,\rm est}^2(t)=(\Delta\psi)^2\;\approx\;\sigma_{\phi,0}^2\Bigl[1+2C\,\epsilon^2 t^2\Bigr],
\end{equation}
with $\sigma_{\phi,0}^2$ the non-relativistic limit. In addition, the dynamics produce a {deterministic} quadratic LO-signal phase drift
\begin{equation}
\psi_{\rm rel}(t)=\gamma t^2,\qquad \gamma=\kappa\,\epsilon^2,
\label{app:eq:drift}
\end{equation}
with geometry-dependent constant $\kappa$ defined in Sec.~\ref{sec:exp_proposal}. If pilot symbols at times $t_p$ are used to correct the following data at $t_p+\Delta t$, the residual is
\begin{align}
\text{ZOH (sample\&hold):}\quad &\delta\phi_{\rm ZOH}=\gamma\bigl[(t_p+\Delta t)^2-t_p^2\bigr]=\gamma(2t_p\Delta t+\Delta t^2),\\
\text{Two-pilot linear predictor:}\quad &\delta\phi_{\rm lin}\approx \tfrac12\psi''(t_p)\,\Delta t^2=\gamma\,\Delta t^2,
\end{align}
so linear prediction cancels the large $2t_p\Delta t$ term \cite{Laudenbach2019Quantum,Marie2017PRA}. The total phase error $\phi=\phi_{\rm est}+\delta\phi_{\rm drift}$ has mean square
\begin{equation}
\sigma_\phi^2(t)=\sigma_{\phi,\rm est}^2(t)+\bigl(\delta\phi_{\rm ZOH/lin}\bigr)^2
\;\approx\;\sigma_{\phi,0}^2\Bigl[1+2C\,\epsilon^2 t^2\Bigr]\;+\;\bigl(\delta\phi_{\rm ZOH/lin}\bigr)^2,
\end{equation}
which, inserted into the small-angle mapping, gives the relativistic addendum to the input-referred excess noise:
\begin{equation}
\Delta\xi_{\rm rel}(t)\;\approx\;\Bigl\{\,2C\,\epsilon^2 t^2\,\sigma_{\phi,0}^2\;+\;\bigl[\delta\phi_{\rm ZOH/lin}\bigr]^2\,\Bigr\}\,\Bigl(V_A+\frac{1}{T}\Bigr).
\end{equation}
This is the same way LLO phase-noise is budgeted in satellite CV-QKD feasibility analyses \cite{Dequal2021,Shao2021PRA,Kish2020QE}.

The secret-key rate in the asymptotic, reverse-reconciled Gaussian model is
\begin{equation}
K_\infty=\beta\,I_{AB}(T,\chi_{\rm tot},V_A)-\chi_{BE}(T,\chi_{\rm tot},V_A),
\end{equation}
with reconciliation efficiency $\beta$. For homodyne detection,
\begin{equation}
I_{AB}=\tfrac12\log_2\!\left(\frac{V+\chi_{\rm tot}}{1+\chi_{\rm tot}}\right),\qquad V=V_A+1,
\end{equation}
and the same expression without the factor $\tfrac12$ for heterodyne \cite{Weedbrook2012RMP}. Since
\begin{equation}
\frac{1}{\ln2}\!\left(\frac{1}{V+\chi_{\rm tot}}-\frac{1}{1+\chi_{\rm tot}}\right)<0\qquad (V>1),
\end{equation}
we have $\partial I_{AB}/\partial\chi_{\rm tot}<0$ for both detection types, while Gaussian extremality and purification imply $\partial\chi_{BE}/\partial\chi_{\rm tot}>0$ at fixed $(T,V)$ \cite{Weedbrook2012RMP}.\;Therefore
\begin{equation}
\frac{\partial K_\infty}{\partial \xi}
=\beta\,\frac{\partial I_{AB}}{\partial \chi_{\rm tot}}
-\frac{\partial \chi_{BE}}{\partial \chi_{\rm tot}} \;<\;0,
\end{equation}
so any positive $\Delta\xi_{\rm rel}(t)$ strictly lowers $K_\infty$ and reduces the operating region where $K_\infty>0$. In finite size, $\xi$ is upper-bounded from data; a larger, time-varying $\xi$ inflates that bound and increases abort probability.

For practice: estimate $\gamma=\kappa\epsilon^2$ by fitting pilot-only sequences (and track slow updates), obtain $C$ from measured pilot/data powers via Eq.~\eqref{app:eq:bhd_rms}, pre-compensate a $\gamma t^2$ phase or use two-pilot linear prediction so $\delta\phi_{\rm ZOH}\!\to\!\delta\phi_{\rm lin}=\gamma\Delta t^2$, raise pilot SNR and shorten the estimation window $t$ to suppress the random bracket, and re-optimise $V_A$ since $\Delta\xi_{\rm rel}\propto (V_A+1/T)$.

\twocolumngrid

\bibliography{references}

\end{document}